# Evidence of Hot Carrier Extraction in Metal Halide Perovskite Solar Cells

S. Sourabh[1], H. Afshari[1], V. R. Whiteside[1], G. E. Eperon[2], R. A. Scheidt[3], T.D. Creason[4], M. Furis[1], A. Kirmani[3], B. Saparov[4], J. M. Luther[3], M. C. Beard[3] and I. R. Sellers[1*]

[1]Department of Physics and Astronomy, University of Oklahoma, Norman, OK 73019
[2]Swift Solar, San Carlos, CA 94070
[3]National Renewable Energy Laboratory, Golden, CO 80401 United States
[4]Department of Chemistry & Biochemistry, University of Oklahoma, Norman, OK 73019

The presence of hot carriers is presented in the operational properties of an (FA,Cs)Pb(I, Br, Cl)$_3$ solar cell at ambient temperatures and under practical solar concentration. At 100 K, clear evidence of hot carriers is observed in both the high energy tail of the photoluminescence spectra and from the appearance of a non-equilibrium photocurrent at higher fluence in light *J-V* measurements. At room temperature, however, the presence of hot carriers in the emission at elevated laser fluence are shown to compete with a gradual red shift in the PL peak energy as photo induced halide segregation begins to occur at higher lattice temperature. The effects of thermionic emission of hot carriers and the presence of a non-equilibrium carrier distribution are also shown to be distinct from simple lattice heating. This results in large unsaturated photocurrents at high powers as the Fermi distribution exceeds that of the heterointerface controlling carrier transport and rectification.

*Corresponding Author: sellers@ou.edu

An increasing global demand for renewable energy and the sustainable implementation of these technologies is challenging the solar research community to develop new materials and increase the power conversion efficiency, while continuing to reduce the operational costs of photovoltaic (PV) systems. Recently, metal halide perovskite solar cells have stimulated tremendous interest due to their facile fabrication and the potential for large area cheap roll-to-roll manufacturing protocols.

Moreover, in less than 10 years these materials have demonstrated solar cell power conversion efficiencies in excess of thin film or multi-crystalline silicon at > 25% [1]. Notably, perovskite based tandems have also recently exceeded the performance of *single junction* GaAs solar cells [2].

While poor stability, low yield, and reproducibility concerns continue to inhibit the large-scale implementation of the perovskites, tremendous progress has been made to stabilize the ABX$_3$ structure with operation of perovskite solar cells now demonstrated in excess of 1000 hours in ambient conditions for both Pb [3] and Pb-Sn [4] based systems.

A significant innovation to improve the stability of perovskite absorbers has been the fine tuning of their chemical compositions to include alloying on both the A-site cation and the X-site halide anion such as (for example) the so-called FAMACs systems [5] or triple halide perovskites under development for tandems [6]. Such alloying, aimed at optimizing the structural tolerance factor, has been shown to be effective in improving materials stability.

While they are compositionally more complex than early perovskite systems such as MAPbI$_3$ or FAPbI$_3$, combinations of double or triple cations and/or halides have been shown to produce more thermodynamically and crystallographically pure structures (low strain, less disordered) that inhibit parasitic or irreversible halide segregation and disorder [7-10]. A significant benefit of these advances is the ability of perovskites to withstand extreme thermal cycling [5, 11, 12] and high temperatures, as well as, to inhibit UV degradation [6].

Moreover, several recent studies have also indicated the presence of long-lived hot carriers in perovskite nanocrystals [13, 14], traps [15], transient absorption measurements [16], luminescence [17, 18], transport [19] along with evidence of a phonon bottleneck [20] and inhibited heat dissipation in these systems [21]. This suggests perovskites may have potential in advanced concept PV [22] such as the hot carrier solar cell [23, 24] as discussed by Li *et al.* [25]. Here, evidence is provided that indicates the exciting hot carrier dynamics observed spectroscopically by the community translates to solar cells based on these materials.

Here, it will be shown that hot carriers can be observed *in the photovoltaic response* of $FA_{0.8}Cs_{0.2}PbI_{2.4}Br_{0.6}Cl_{0.02}$ solar cells under practical operational conditions further supporting the potential of these systems for hot carrier devices.

The details of the devices studied here are as follows: the solar cells consist of a 200 nm spin-coated $FA_{0.8}Cs_{0.2}PbI_{2.4}Br_{0.6}Cl_{0.02}$ absorber layer integrated into a device structure comprising an ITO front contact, PolyTPD (poly(N,N-bis- 4-butylphenyl-N,N-bisphenyl)-benzidine) hole transport layer with a PFN-Br (poly[(9,9-bis(3-(N,N-dimethylamino)propyl)-2,7-fluorene)-alt-2,7-(9,9-dioctylfluorene)]dibromide) interface layer. The back of the device includes, $ZTO/SnO_2$ and $C_{60}$ electron transport layers with a LiF interfacial layer, followed by an ITO back-contact. The devices were completed with a 50 nm $TiO_2/Al_2O_3$-based multilayer encapsulant to improve the atmospheric stability. A schematic of the full structure and materials information is provided in the supplementary information (**SI-1**).

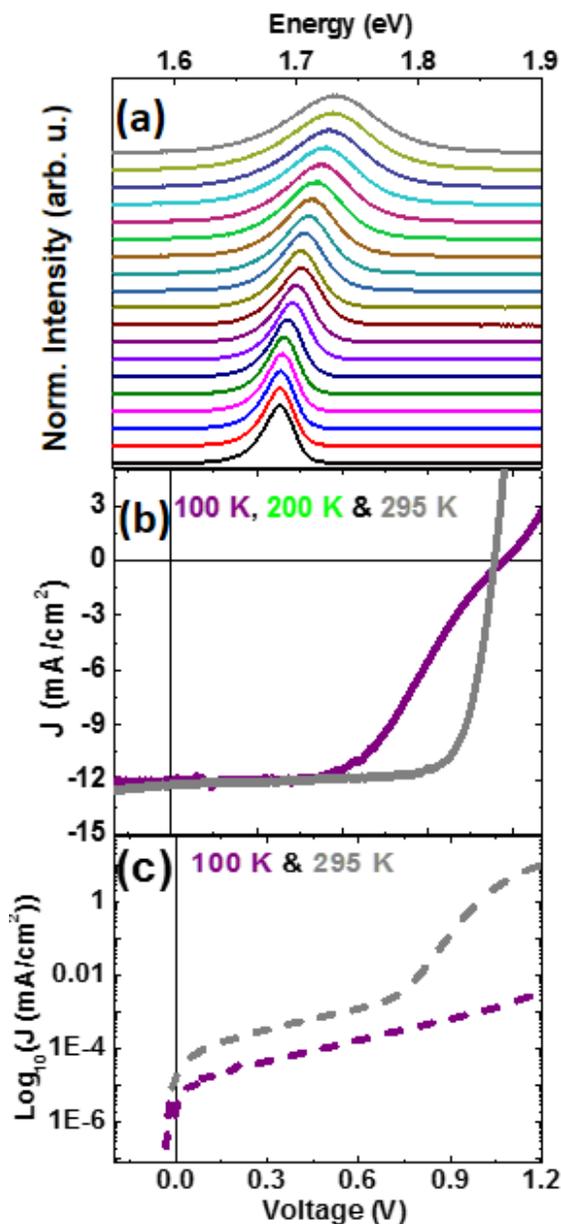

Figure 1: (a) Normalized waterfall plot of temperature dependent photoluminescence from 4 K to 295 K. (b) Comparison of the low power (1-sun AM 1.5G equivalent) monochromatic (532 nm excitation) current density – voltage responses at 100 K (purple), 200 K (green), and 295 K (gray) along with their respective dark JVs (same color code).

Initial assessment of the *J-V* responses of the devices were tested in the dark and under 1-sun

AM 1.5G illumination using a Newport class ABA solar simulator referenced to a calibrated silicon solar cell (these data are provided in the supplementary information, **SI-2)**. To assess the presence of hot carriers and the role of halide segregation and materials decomposition simultaneous current-density voltage (*J-V*) and photoluminescence (PL) measurements were performed from 4 K to 295 K in a Janis closed cycle cryostat system. A 532 nm laser at power densities ranging from 20 mW/cm$^2$ to 8000 mW/cm$^2$ (0.2 to 80 suns equivalent) was used as the excitation source.

Figure 1 (a) shows the temperature dependent (TD) PL spectra for a representative $FA_{0.8}Cs_{0.2}PbI_{2.4}Br_{0.6}Cl_{0.02}$ solar cell from 4 K to 295 K. With increasing temperature, the well-known blue shift in the perovskite emission energy is observed consistent with the s and p-orbital nature of perovskite's valence and conduction bands, respectively [26]. While there is no convincing evidence of a phase transition in the temperature range studied (see **SI-3**), phase evolutions cannot be totally discounted.

In addition to this increase in the PL energy with temperature, a noticeable broadening of the PL spectra is also evident, which is representative of the strong and well-studied Fröhlich interaction and the strong polar nature of these materials [21, 27-29]. These properties were confirmed for the system under investigation via analysis of the phonon induced PL broadening [30-32] which demonstrated the strong contribution of LO-phonons in the linewidth of the PL. This analysis is available in the supplementary information (**SI-3**) [20, 30, 33].

The monochromatic low power (532 nm/1 sun-AM 1.5G equivalent) *J-V* characteristics measured simultaneously with PL at 100 K, 200 K, and 295 K are shown in Figure 1(b), with the associated dark *J-V* shown on a log scale in Figure 1(c). The full set of *J-V* spectra from 4 K to 295 K is given in the supplementary information, Figure **SI-4**. At 100 K (solid purple line) the fill factor (FF) of the *J-V* is dominated by a strong s-shaped inflection consistent with the presence of a parasitic barrier[34] to minority carrier extraction. This is also reflected in the low dark current and high resistance to carrier transport at 100 K.

Such behavior has been observed previously in the perovskites and has been attributed to the presence of a non-ideal interface and/or barriers created within the structure due to different temperature coefficient of the constituent materials of the solar cells that induce carrier localization [5, 35, 36]. This localization has been shown to be correlated to the PL intensity, which is strong in the presence of inhibited carrier extraction and reduces as the carrier extraction increases at elevated temperatures [5, 36].

The presence of this barrier and the extraction of carriers has been shown to be driven by the competition between thermionic emission and the carrier generation rate, along with the voltage dependence of the heterointerfaces. This is further supported by the presence of negative differential resistance in forward bias in the majority carrier regime (dark and/or strong applied voltage) [5, 11].

As the temperature is increased to 200 K (solid green line – Fig. 1(b)) and further to 295 K (solid gray line – Fig. 1(b)) the fill factor and series resistance increase and decrease, respectively. Simultaneously the dark *J-V* also improve (Fig. 1(c)). These data indicate that the band offsets and/or the thermal energy of the carriers improve at higher temperatures (> 200 K) facilitating carrier extraction, and that the device structure and operation are well matched to the optimal conditions for PV operation under 1-sun conditions.

To further assess the role of lattice temperature on the operating conditions of the solar cell, and more subtly, the role and presence of hot carriers in the system, simultaneous power dependent (PD) PL and *J-V* were performed at 100 K, 200 K, and 295 K. Figure(s) 2(a), 2(b), and 2(c) show the PD PL at 100 K, 200 K, and 295 K, respectively.

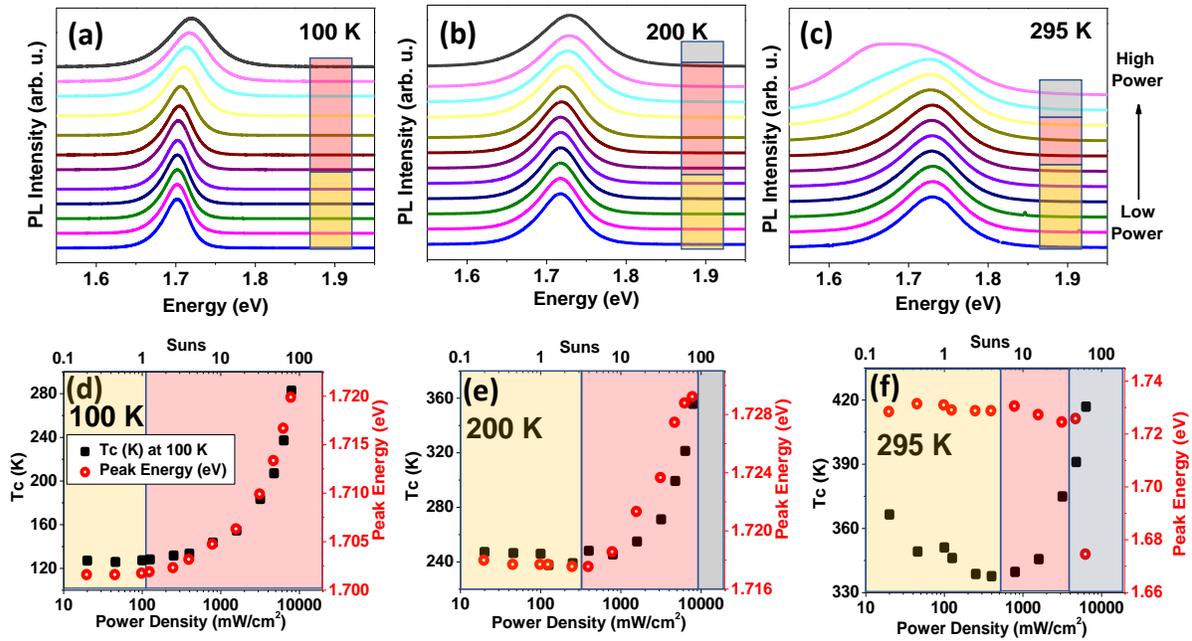

Figure 2: Power dependent photoluminescence (PD PL) spectra for: (a) 100 K, (b) 200 K, and (c) 295 K excited at 532 nm from 20 mW/cm² to 8000 mW/cm² (295 K is from 20 mW/cm² to 6400 mW/cm² ) A comparison of the carrier temperature ($T_c$) (solid black squares) and the peak energy (open red circles from the data shown in (a) 100 K, (b) 200 K, and (c) 295 K, are shown in (d), (e), and (f), respectively. The yellow shaded region represents the equilibrium carrier temperature, while the pink region represents the regime in which non-equilibrium hot carriers are present. At higher temperatures halide segregation becomes prevalent shown in the gray shaded region. These regimes are also indicated by the color-coded boxes in (a), (b), and (c) on the high energy tail of the PL spectra.

At 100 K as the power is increased from 20 mW/cm² to 8000 mW/cm² (0.2-sun AM 1.5 G to 80 suns equivalent) the PL blue shifts and broadens indicating an increase in energy in the system. This is attributed to the increased carrier concentration (band filling) though other effects such as exciton-exciton interactions can't be discounted. Despite the presence of LO-phonon mediated broadening beginning to screen the PL shift at 200 K, a slight shift is also evident at this intermediate temperature. This is more clearly visualized in Figure(s) 2(d) and 2(e), which show the peak energy extracted as a function of power from the PL (open red circles) at 100 K and 200 K, respectively.

At 295 K the integrated intensity of the low power PL is **3** orders of magnitude lower than that at 100 K (the 200 K PL is **2** orders lower than that at 100 K – see **SI-5**) as a result of the efficient exciton dissociation at elevated temperatures [37]. As the intensity of the laser fluence is increased at 295 K, little evidence for hot carriers is observed in the PL response of the device (Figure(s) 2(c) and 2(f)). Notably, there is no observed shift in peak energy either from the PL spectra (Figure 2(c)) or from the extracted peak energy, Figure 2(f)). This is attributed to the combination of the relatively low PL intensity [31, 32], reversible halide segregation at higher fluences (pink), and the effects of ionic induced defect formation under illumination [38].

At excitation powers in excess of 3100

mW/cm$^2$, clear evidence of halide segregation is observed in the system (Figure 2(c)), which is supported by a large reduction in the peak energy at high powers (Figure 2(f)), and that negatively affects the device performance – as discussed below with respect to Figure 5.

Figure(s) 2(d), 2(e), and 2(f) show a comparison of the peak energy extracted from the PD PL at 100 K, 200 K, and 295 K compared to the carrier temperature ($T_C$) extracted from the slope of the CW PL using a generalized form of the Planck equation. Such analysis, which has been developed by the III-V PV community [39-42] was applied recently in several forms to study perovskites in transient absorption [18, 20] and in steady state CW PL measurements [17, 18, 43].

Here, a simple linear fit of the high energy slope is used is extract the carrier temperature – while also subtracting the contribution of [18, 44] the phonon broadening (see **SI-3** and **SI-6**) to garner the *qualitative trend* in carrier heating [45]. This is then supported by the simultaneous PD *J-V* (Figure(s) 4 and 5) to elucidate the presence of hot carriers. A further caveat is that such analysis *should* only be performed upon PL that is not affected by decomposition/ halide segregation [31, 46, 47]. This prevents erroneous results that invalidate the use of the generalized Planck's equation as described previously [46].

Figure 2(d) shows a comparison of $T_C$ and the peak energy at 100 K extracted from the data in Figure 2(a). At excitation powers below 100 mW/cm$^2$, the carrier temperature is independent of power and the system is considered at equilibrium (yellow shaded region) and therefore both the photogenerated carriers and lattice have a global temperature of ~ 100 K [44, 48].

However, as the excitation power is increased from 100 mW/cm$^2$ to 8000 mW/cm$^2$, $T_C$ increases (open black circles) from 100 K to 280 K (or a $\Delta T_C$ of ~ 180 K at maximum power). This increase in temperature can be attributed to either lattice heating *or* the presence of hot carriers in the device. As we discuss below in the *J-V* measurements, we attribute this increase in temperature partially to the presence of hot carriers that dominate predominately at lower temperatures. Moreover, whilst $T_C$ increases, there is also a simultaneous increase on the peak PL energy (open red squares) from ~ 1.7 eV to 1.72 eV.

At 200 K, a similar behavior is observed with $T_C$ increasing from an equilibrium temperature of ~ 200 K below 400 mW/cm$^2$ (yellow shaded region) to a $T_C$ of approximately 360 K ($\Delta T_C$ of ~ 160 K – pink shaded region) at ~ 6300 mW/cm$^2$. At 200 K, a combination of the increased thermal energy ($k_B T$) and the energy pumped into the system via laser excitation also results in a threshold for halide segregation, depicted by the gray shaded region in Figure 2(b) at $P > 6300$ mW/cm$^2$.

In this regime decomposition of the perovskite and halide segregation broaden the PL and result in a reduction of the peak PL energy [10, 49]. While the peak PL energy shown as a function of power in Figure 2(e) once again blue shifts from ~ 1.718 eV to ~ 1.729 eV at increasing power, unlike the 100 K data, the PL energy begins to *decrease* as halide segregation results in PbI$_3$ rich regions [8, 49] that lower the energy of the PL, which is dominated by these low energy inclusions.

This is more clearly evident in Figure 2(f) which shows a comparison of $T_C$ and the peak energy at a lattice temperature of 295 K. Again, at 295 K the phonon mediated broadening of the PL screens any shift in the peak PL energy with increasing fluence, and at higher (pink and gray) powers the PL transitions from an equilibrium regime (yellow region) to that dominated by halide segregation (gray region) crossing a screened hot carrier regime (pink region). This is further supported by the line shape of the PL evident for this temperature range and power regime shown in Figure 2(c), and the reduction of the peak PL energy evident in Figure 2(f).

Furthermore, since decomposition changes the PL linewidth with increasing powers extraction of $T_C$ at 295 K is compromised, and evidence of hot carriers at 295 K is best supported by the PD *J-V* at the same lattice temperature, discussed further below. Previous research has probed the energy of activation for halide ion photo segregation for many species of perovskites and found that value to be approximately 80-600 meV [50-53]. Therefore, we can assume that while the energy of the hot carriers in the system exceeds this value, they also serve to accelerate the photo segregation process, particularly at higher temperature (295 K).

While the preceding discussion indicates that the system experiences carrier heating under high fluence, the nature of this heat and the role of increased lattice heating versus effects due to photo-excited non-equilibrium "hot carriers" requires further consideration. In Figure 2(d, e, and f) the extracted $T_C$ increases in sync with a blue shift in the peak PL energy indicating the system is heating, without elucidating information on the nature of the increased thermal energy within the system. When comparing the temperature dependence of the PL with the shift in the PL peak energy from power dependence measurements at the equilibrium lattice temperature of 100 K and 200 K (shown in supplementary information, **SI-7**), the excitation energy from the laser apparently increases the temperature by ~ 125 K (from 100 K to ~ 225 K) and 75 K (200 K to ~ 275 K), respectively.

While lattice heating due to the low thermal conductivity of perovskites appears a simple conclusion to the thermal effects observed in Figure 2, inspection of the simultaneously measured *J-V* responses (with the PL in Figure(s) 2(a – c)) indicate that the observed effects are subtler than those derived by simple inhibited heat dissipation arguments in these measurements.

Figure 1(b) indicates that increasing the temperature of the device at 1-sun AM 1.5G equivalent excitation improves the FF and the PV response of the solar cell due to increased carrier extraction in the heterostructure architecture. When considering the *J-V* response of the device as a function of laser fluence at 100 K, 200 K, and 295 K, as shown in Figure 3(a), (c), and (e) respectively, the *J-V* response shows somewhat different behavior to that of increasing (equilibrium) lattice temperature (Figure 1(b)), where the FF improves as a function of temperature.

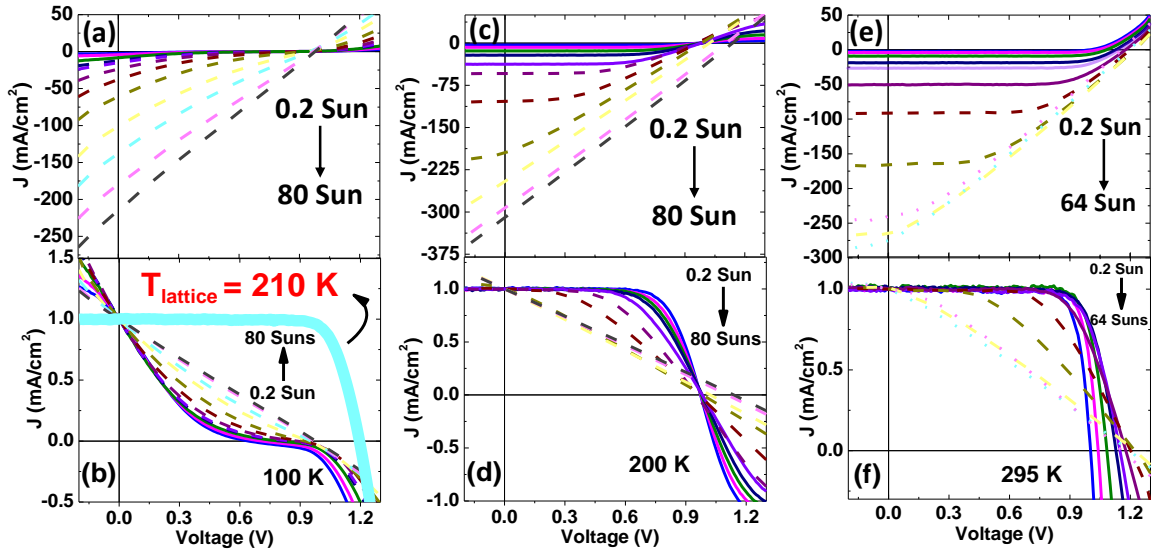

Figure 3: Power dependent *J-V* at (a) 100 K, (c) 200 K and (e) 295 K respectively from low power to high power. The same data are shown normalized to *Jsc* in (b), (d), and (f), respectively, used to illustrate hot carrier related effects and the associated Fermi-distribution. The dashed, solid and dotted lines represent hot carrier, equilibrium and phase segregation regime respectively. The solid cyan curve in (b) represent the *J-V* at lattice temperature of 210 K and compared to equivalent temperature of 207 K in dotted cyan at 47 Sun.

In Figure 3(a) the power dependent *J-V* response is shown at 100 K. In Figure 3(b) the same data is normalized with respect to *Jsc (V=0)* to assess the role of increasing power upon the shape of the *J-V* response following the analysis of Dimmock *et al*. [54]. These *J-V* responses (and all those in Figure(s) 3) are plotted with respect to the three regimes described in Figure 2 for the associated PL. The solid lines in the *J-V* data represent the equilibrium temperature region, while the dashed and dotted represent the hot carrier and phase segregation regimes, respectively.

In Figure 3(a) the *J-V* data at 100 K once more displays the non-ideality of the system at lower temperatures [3, 36]. This is amplified in the normalized *J-V* data at 100 K shown in Figure 3(b). Moreover, as the laser fluence is increased the rectification of the diode is removed, the FF decreases, and the device becomes more resistive with a large increase in $J_{sc}$ (Figure 3(a)). The magnitude of $J_{sc}$ scales with the laser fluence with an absence of saturation of the photogenerated carrier collection that would reflect the absorbed photon flux and infinite shunt resistance of conventional PV operation (Figure 3(b)).

The transition to a linear resistive-like *J-V* response observed at higher (dashed lines) powers in Figure 3(a) and 3(b) is indicative of the presence of a non-equilibrium hot carrier population, the tail of which exceeds the parasitic barrier in the system. The shape of the *J-V* at V < $V_{oc}$ then reflects the shape of the Fermi distribution with increasing reverse bias [54].

Further evidence that the *J-V* is dominated by non-equilibrium carrier extraction (hot photocurrent) and not simple lattice heating is provided by the significant difference in the *equilibrium J-V* response at 210 K with respect to the 100 K PD *J-V* at ~ 47 Suns (i.e., the PD *J-V* with an equivalent $T_C$ ~ 207 K). These data are compared directly in Figure 3(b). The *equilibrium J-V* at 210 K (solid cyan line) shows a typical photovoltaic response with a $J_{sc}$ of ~ 12 mA/cm². While the high-power PD *J-V* with an effective temperature of 207 K (dashed cyan line) displays a photoconductive-like response and a large $J_{sc}$ of ~ 320 mA/cm². These data are shown normalized and independently of the other *J-V* spectra for additional clarity in the supplementary information, **SI-8**.

This comparison (Figure 3(b)) is *critical* in demonstrating that the power dependent *J-V* data reflect a system in which non-equilibrium carriers exist in a structure that remains at 100 K, despite the high power excitation the system experiences. This is evident since the presence of parasitic barriers, non-ideal interfaces, and/or inhibited conductivity experienced by the photogenerated carriers at *low temperatures* (Figure 1) is still evident (poor fill factor) under high laser excitation, in Figure 3(b). This suggests the *lattice* temperature remains at ~ 100 K, while the carrier temperature is significantly greater (~ 207 K).

If the lattice was heated due to the high power laser illumination it would result in a *J-V* response consistent with the temperature data at ~ 200 K (solid cyan – Figure 3(b)) – this is clearly not the case. The presence of hot carriers is also supported by the extraction and *J-V* parameters at high power discussed further below.

The behavior of the devices at higher (dashed lines) power and low temperature (100 K) is consistent with the generation and subsequent extraction of a hot carrier distribution, the tail of which exceeds the parasitic barrier and/or heterointerface [54]. As the bias is further increased in the negative direction, the extraction of the hot carriers via thermionic emission begins to compete with direct tunneling, as the potential barrier narrows. In this regime the slope of the *J-V* reflects the resistance to minority carrier extraction from the carrier distribution, in general, and the limited role of the heterostructure in producing diode like rectification, which is removed at high laser fluence, resulting in photoconductive-like behavior more ohmic in nature while losing rectification.

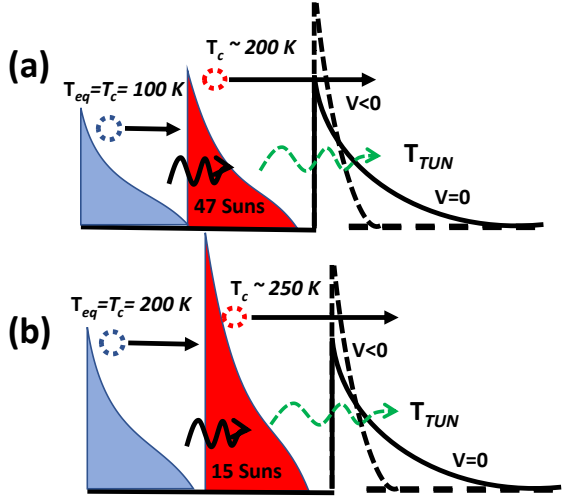

Figure 4: (a) Schematic representation of the equilibrium carrier temperature ($T_{eq}$) at 100 K (blue) and the non-equilibrium carrier temperature ($T_C$) induced at 100 K when excited with 4700 mW/cm$^2$ (~47x sun equivalent) with a $T_C \sim 200$ K (red). These are shown with respect to a parasitic barrier that mimics the behavior seen experimentally at $V = 0$ (solid), and $V < 0$ (dashed) and the relative contribution the thermionic emission (solid black arrow) and tunneling ($R_{TUN}$ – dotted green arrow), respectively. This is illustration is repeated for $T_{eq} = 200$ K in (b).

This phenomenological protocol is shown schematically in Figure 4(a), which illustrates the electron Fermi distribution at 100 K (blue) and 200 K (red) to the left of a potential barrier with an applied bias of $V = 0$ V (solid line) and in reverse bias (dotted line). At 100 K, the carrier distribution experiences a large potential barrier until significant reverse bias is applied when breakdown would occur. In the case of the hot carrier distribution (red – 47 suns), there is a significant tail above the barrier increasing the thermionic emission, $T_{TE}$ resulting in greater carrier extraction and the absence of a saturated $J_{sc}$. As the reverse bias narrows the barrier $T_{TE}$ competes with direct tunneling, $T_{TUN}$.

As the lattice temperature is increased from 100 K to 200 K carrier transport through the device is facilitated by increased thermal energy, along with a subsequent reduction of the contribution of parasitic resistances and non-ideal heterointerfaces within the solar cell architecture. Despite the improved PV functionality of the solar cells at 200 K, evidence of the parasitic heterointerfaces *is* evident in a high series resistance at low powers ($P < 250$ mW/cm$^2$) – Figure 1(b). Evidence of a photoconductive response and the presence of hot carriers is also apparent in Figure 3(d) at $P > 1000$ mW/cm$^2$ (dashed lines) and the related PL.

Figure 4(b) shows the 200 K case schematically with the higher thermal energy reducing the effect of the barrier to carrier transport relative to that at 100 K. Once again, thermionic emission and direct transport of hot carriers across the heterojunction serves to reduce the rectifying properties of the solar cell by limiting the effect of the electric field in modulating carriers at the interface. While the nature of the hot carrier type (electron and/or hole) creating this effect, or indeed which interface is responsible for modulating the carrier transport is not fully understood, the role of hot carriers in the structure *is* clear. These measurements, confirm that non-equilibrium carriers are sustained in perovskites at high fluences, but more importantly *here it is also shown this occurs at levels achievable in practical concentrator PV systems* [80 Suns].

Indeed, similar effects to the ones presented here under monochromatic excitation have been seen under concentrated solar illumination in analogous FAMACs devices [6] (see **SI-9**), *further supporting the potential of perovskites for practical hot carrier solar cells.*

An example and a schematic representation of potential operation at a representative parasitic interface is shown in the idealized energy band diagram given in the supplementary information, **SI-10**. However, further work is required to elucidate the exact interface responsible in the device under study.

Although hot carriers are extracted by thermionic emission at high fluences, these carriers cool in the non-optimized (for hot carrier extraction) contacts which set the voltages below those expected for ideal HCSC operation [24, 55]. These data do, however, suggest such operation is feasible with device architectures designed to optimize such operation with the implementation of degenerate contact layers that match the hot carrier distribution, provided there are heterointerface configurations that support rectification of carrier populations with energy greater than the perovskite band gap.

At 295 K the conditions for high efficiency operation of perovskite solar cells under 1-sun conditions is met. [3]. The majority and minority carrier transport is then dominated by diffusion across the nominally intrinsic perovskite absorber, and the extraction and rectification of carriers at high field regions and at the heterointerfaces between the perovskite and the electron and hole transporting layers [56].

Interestingly, however, evidence for hot carriers is also observed in the *J-V* response at ambient temperatures as shown in Figure(s) 3(e) and 3(f). Specifically, as the power is increased above 600 mW/cm$^2$, which is ~ 6 suns equivalent, the parasitic effects of hot carriers in these solar cells becomes evident in the loss of strong rectification of the heterostructure diode. That is, the FF is reduced and the shape of the *J-V* begins to mimic the carrier distribution below $V_{oc}$ [54].

While the devices presented display clear evidence of hot carrier effects via the extraction of *hot photocurrent,* they do not show increased efficiency. Furthermore this is not the realization of an operational hot carrier solar cell (HCSC) [24], since the structures are not designed to extract hot carriers selectively and the operation of these solar cells – particularly

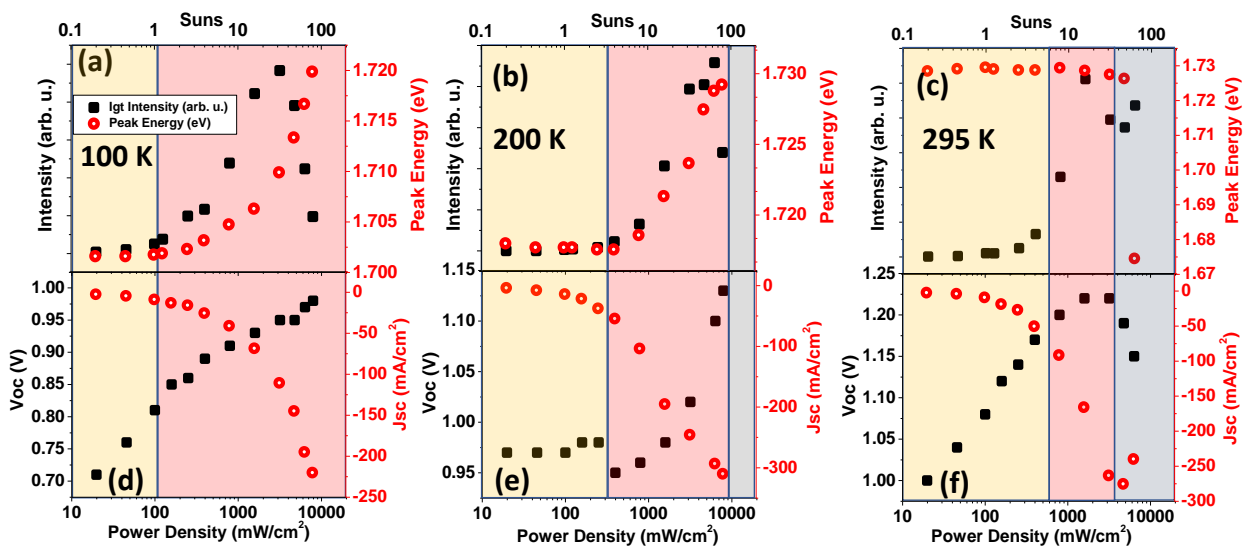

Figure 5: A comparison of the power dependent integrated PL intensity and peak PL energy at (a) 100 K, (b) 200 K, and (c) 295 K, respectively. In each case these can be associated to the lower panels that compare the simultaneously measured $V_{oc}$ and $J_{sc}$ at (d) 100 K, (e) 200 K, and (f) 295 K, respectively. The shaded regions represent the regimes at equilibrium (yellow), in the non-equilibrium hot carrier regime (pink), and at high power where halide segregation is evident (gray).

$V_{oc}$ – remains set by the absorber band gap.

Further evidence to support the presence of hot carriers, thermionic emission of a non-equilibrium carrier population, and their role on solar cell operation are provided by comparing the PV parameters extracted from the PD *J-V* responses to those of the PL. Figure 5 shows a compilation of the peak PL energy and the integrated emission, along with the extracted $J_{sc}$ and $V_{oc}$ as a function of power at 100 K, 200 K, and 295 K.

The power regimes are once again (consistent with Figure 2), color coded to represent regimes in which the system is deemed in equilibrium (shaded yellow), in the hot carrier regime (shaded pink), or where the high power and temperature have induced halide segregation (shaded gray).

Figures 5(a) and 5(d) compare the power dependence of the PL properties and PV parameters at 100 K, respectively. With increasing laser fluence, the PL intensity increases as expected. Behavior that is enhanced by the localization of inhibited transport of carriers by the presence of parasitic barriers [5, 36].

However, at $P > 1600$ mW/cm$^2$ (16 Suns), a significant drop in the PL intensity is evident whilst the peak PL energy continues to blue shift, as the carrier temperature increases towards $T_C \sim 200$ K. This drop in PL intensity is attributed to the thermionic emission of hot carriers and the increased extraction of the photogenerated carrier distribution. This is supported by a simultaneous increase in $J_{sc}$ (open red circles – Figure 5(d)) at high powers, and the exponential dependence of $J_{sc}$ at $P > 1600$ mW/cm$^2$. This is further support for the role of the non-equilibrium Fermi-distribution, the tail of which exceeds the parasitic heterointerface/barrier.

The absence of parasitic effects and carrier cooling in the contacts is apparent when considering the effect of the increasing fluence on the $V_{oc}$ (open black circles) at 100 K, shown in Figure 5(d). At lower power ($P < 1600$ mW/cm$^2$) $V_{oc}$ increases from 0.7 V to ~ 0.9 V which represents the steady increase in the quasi-fermi ($\Delta q_{fermi}$) levels and typical behavior of a PV device under increasing illumination. At maximum fluence ($P \sim 8000$ mW/cm$^2$ – 80 sun equivalent), $V_{oc}$ increases to ~ 1 V which is consistent with the band gap of the perovskite absorber and the carrier concentration generated ($\Delta q_{fermi}$).

The correlation between non-equilibrium carrier generation and thermionic extraction, increasing $V_{oc}$, and simultaneous quenching of PL at the highest powers is also evident at 200 K – shown in Figure(s) 5(b) and 5(e). Notably, there is a slight drop in Voc at the onset of non-equilibrium carrier generation that coincides with a loss of rectification. Once again, the competition with the thermal energy of the photogenerated carriers and the lowering of the potential barriers results in an increasing $J_{sc}$ and reduced PL intensity at the very highest power. At 295 K analysis of the optical and optoelectronic behavior allows a distinction in the effects due to non-equilibrium hot carriers (observed at lower temperatures) and effects due to decomposition of the perovskite, more specifically halide segregation.

In Figure 5(c) the laser fluence is observed to have little effect upon the peak PL energy, with a slight reduction in this energy observed at $P > 800$ mW/cm$^2$, which is correlated with an increase in the integrated PL intensity (solid black squares). When comparing these effects with the PL spectra at 295 K in Figure 2(c) it is clear the emission is derived from considerable broadening of the PL and the appearance of a low energy feature attributed to the formation of iodine rich regions in the perovskite [9, 10]. These halide inclusions begin to dominate the PL spectrum and reduce the "effective" band gap for emission evident at high fluence in Figure 5(c) (open red circles).

This transition from equilibrium behavior (yellow shaded) to a regime where material decomposition begins to dominate (gray shaded) is also reflected in the PV parameters at 295 K at higher fluence (see Figure 5(f)). Specifically, at lower fluences (yellow, pink) there is a simultaneous increase in $V_{oc}$ and $J_{sc}$ with increased photon flux – as expected in a PV device. However, as the power exceeds ~ 3100 mW/cm$^2$, the effects of halide segregation become visible in the PL spectra, and a simultaneous reduction in both $V_{oc}$ and $J_{sc}$ occurs reflecting the increased role of defects and non-radiative channels as halide segregation proceeds under the highest laser fluences (gray) at 295 K.

While the degradation of the perovskite film complicates/limits the assessment of hot carriers at the highest powers (shaded gray) at ambient conditions here, the presence of a regime dominated by halide segregation and the known photo induced degradation (shaded gray) of perovskites independent of the regime in which evidence of non-equilibrium carriers (shaded pink) are present provides strong support for the conclusions of this work, since these regimes are clearly distinguished. This is facilitated by the stability of $FA_{0.8}Cs_{0.2}PbI_{2.4}Br_{0.6}Cl_{0.02}$ under investigation that allows high power excitation and thermal cycling without irreversible degradation of the devices [38].

In summary, evidence of hot carrier effects are presented in a $FA_{0.8}Cs_{0.2}PbI_{2.4}Br_{0.6}Cl_{0.02}$ solar cell at a range of lattice temperatures, effects which are independent and distinct to those of halide segregation or degradation of devices under high fluence excitation in excess of 50 suns equivalent. The effects of thermionic emission of hot carriers and the presence of a non-equilibrium carrier distribution are also shown to be distinct from simple lattice heating. This results in large unsaturated photocurrents at high powers as the Fermi distribution exceeds that of the heterointerface controlling carrier transport and rectification.

Although these measurements support previous spectroscopic measurements that demonstrate inhibited hot carrier thermalization in perovskites, and hot carrier effects are observed in the operation of the solar cell, enhanced power conversion efficiency and demonstration of hot carrier solar cell operation is not demonstrated. While there is evidence of hot carrier extraction, the carriers cool in the contacts resulting in a $V_{oc}$ that is determined by the band gap of the perovskite absorber rather than the hot carrier distribution.

Notably, this is not *a priori* a limitation towards a perovskite based HCSC; however, due to the robustness of the overall design it has proven to be suitable for proof of principle hot carrier effects. To achieve the realization of a perovskite based HCSC would require an architecture designed specifically for energy selective extraction of the hot carriers and contacts that are degenerate with the hot carrier distribution for the proposed enhanced $V_{oc}$.

This work is funded through the Department of Energy EPSCoR Program and the Office of Basic Energy Sciences, Materials Science and Energy Division under Award No. DE- SC0019384. T.D.C and B.S. acknowledge the support from the U.S. Department of Energy, Office of Science, Office of Basic Energy Sciences, under Award No. DE-SC0021158. This work was authored in part by the National Renewable Energy Laboratory, operated by Alliance for Sustainable Energy, LLC, for the U.S. Department of Energy (DOE) under Contract No. DE-AC36-08GO28308. The devices were constructed under support from the Operational Energy Capability Improvement Fund (OECIF) of the Department of Defense (DOD). The views expressed in the article do not necessarily represent the views of the DOE or the U.S. Government.

**Supporting Information (SI)** includes the following parts: For a full description of the sample fabrication and photovoltaic characterization see:
https://pubs.acs.org/doi/10.1021/acsenergylett.3c00551?goto=supporting-info

Device offsets. Temperature dependent JV from 80 K - 300 K. Photoluminescence linewidth broadening analysis. Temperature dependent photoluminescence, transmission/absorption, and monochromatic JV from 4 K - 295 K with peak energy analysis. Generalized Planck analysis. Power dependent JVs at 300 K. Phenomenological representation of a barrier within band diagram.

REFERENCES
1. Min, H., et al., *Perovskite solar cells with atomically coherent interlayers on SnO2 electrodes.* Nature, 2021. **598**(7881): p. 444-450.
2. Jošt, M., et al., *Perovskite/CIGS Tandem Solar Cells: From Certified 24.2% toward 30% and Beyond.* ACS Energy Letters, 2022. **7**(4): p. 1298-1307.
3. Christians, J.A., et al., *Tailored interfaces of unencapsulated perovskite solar cells for >1,000 hour operational stability.* Nature Energy, 2018. **3**(1): p. 68-74.
4. Tong, J., et al., *Carrier control in Sn–Pb perovskites via 2D cation engineering for all-perovskite tandem solar cells with improved efficiency and stability.* Nature Energy, 2022. **7**(7): p. 642-651.
5. Brown, C.R., et al., *Potential of High-Stability Perovskite Solar Cells for Low-Intensity–Low-Temperature (LILT) Outer Planetary Space Missions.* ACS Applied Energy Materials, 2019. **2**(1): p. 814-821.
6. Xu, J., et al., *Triple-halide wide–band gap perovskites with suppressed phase segregation for efficient tandems.* Science, 2020. **367**(6482): p. 1097-1104.
7. Bush, K.A., et al., *Compositional Engineering for Efficient Wide Band Gap Perovskites with Improved Stability to Photoinduced Phase Segregation.* ACS Energy Letters, 2018. **3**(2): p. 428-435.
8. Knight, A.J., et al., *Halide Segregation in Mixed-Halide Perovskites: Influence of A-Site Cations.* ACS Energy Letters, 2021. **6**(2): p. 799-808.
9. Mundt, L.E., et al., *Mixing Matters: Nanoscale Heterogeneity and Stability in Metal Halide Perovskite Solar Cells.* ACS Energy Letters, 2022. **7**(1): p. 471-480.
10. Pavlovetc, I.M., et al., *Suppressing Cation Migration in Triple-Cation Lead Halide Perovskites.* ACS Energy Letters, 2020. **5**(9): p. 2802-2810.
11. Afshari, H., *Radiation tolerance, high temperature stability, and self-healing of triple halide perovskite solar cells*, in *49th IEEE Photovoltaic Specialists Conference (PVSC 49)*. 2022: Philadelphia.
12. Afshari, H., et al., *FACsPb Triple Halide Perovskite Solar Cells with Thermal Operation over 200 °C.* ACS Energy Letters, 2023: p. 2408-2413.
13. Li, M., et al., *Slow cooling and highly efficient extraction of hot carriers in colloidal perovskite nanocrystals.* Nature Communications, 2017. **8**(1): p. 14350.
14. Hopper, T.R., et al., *Hot Carrier Dynamics in Perovskite Nanocrystal Solids: Role of the Cold Carriers, Nanoconfinement, and the Surface.* Nano Letters, 2020. **20**(4): p. 2271-2278.
15. Righetto, M., et al., *Hot carriers perspective on the nature of traps in perovskites.* Nature Communications, 2020. **11**(1): p. 2712.
16. Sum, T.C., et al., *Spectral Features and Charge Dynamics of Lead Halide Perovskites: Origins and Interpretations.* ACCOUNTS OF CHEMICAL RESEARCH, 2016. **49**(2): p. 294-302.
17. Fang, H.-H., et al., *Long-lived hot-carrier light emission and large blue shift in formamidinium tin triiodide perovskites.* Nature Communications, 2018. **9**(1): p. 243.
18. Lim, S.S., et al., *Hot carrier extraction in CH3NH3PbI3 unveiled by pump-push-probe spectroscopy.* Science Advances. **5**(11): p. eaax3620.
19. Guo, Z., et al., *Long-range hot-carrier transport in hybrid perovskites visualized by ultrafast microscopy.* Science, 2017. **356**(6333): p. 59-62.
20. Yang, Y., et al., *Observation of a hot-phonon bottleneck in lead-iodide perovskites.* Nature

Photonics, 2016. **10**(1): p. 53-59.
21. Sourabh, S., et al., *Hot carrier redistribution, electron-phonon interaction, and their role in carrier relaxation in thin film metal-halide perovskites.* Physical Review Materials, 2021. **5**(9): p. 095402.
22. Green, M.A., *Third generation photovoltaics: Advanced Solar Energy Conversion.* Springer Series in Photonics, ed. B.M. T. Kamiya, H. Venghaus, Y. Yamamoto 2003, The Netherlands: Springer. 162.
23. König, D., et al., *Hot carrier solar cells: Principles, materials and design.* Physica E: Low-dimensional Systems and Nanostructures, 2010. **42**(10): p. 2862-2866.
24. Ross, R.T. and A.J. Nozik, *Efficiency of hot-carrier solar energy converters.* Journal of Applied Physics, 1982. **53**(5): p. 3813-3818.
25. Li, M., et al., *Slow Hot-Carrier Cooling in Halide Perovskites: Prospects for Hot-Carrier Solar Cells.* Advanced Materials, 2019. **31**(47): p. 1802486.
26. Filip, M.R., C. Verdi, and F. Giustino, *GW Band Structures and Carrier Effective Masses of CH3NH3PbI3 and Hypothetical Perovskites of the Type APbI3: A = NH4, PH4, AsH4, and SbH4.* The Journal of Physical Chemistry C, 2015. **119**(45): p. 25209-25219.
27. Ghosh, D., et al., *Polarons in Halide Perovskites: A Perspective.* The Journal of Physical Chemistry Letters, 2020. **11**(9): p. 3271-3286.
28. Herz, L.M., *Charge-Carrier Dynamics in Organic-Inorganic Metal Halide Perovskites.* Annual Review of Physical Chemistry, 2016. **67**(1): p. 65-89.
29. Srimath Kandada, A.R. and C. Silva, *Exciton Polarons in Two-Dimensional Hybrid Metal-Halide Perovskites.* The Journal of Physical Chemistry Letters, 2020. **11**(9): p. 3173-3184.
30. D'Innocenzo, V., et al., *Excitons versus free charges in organo-lead tri-halide perovskites.* Nature Communications, 2014. **5**(1): p. 3586.
31. Esmaielpour, H., et al., *Enhanced hot electron lifetimes in quantum wells with inhibited phonon coupling.* Scientific Reports, 2018. **8**(1): p. 12473.
32. Wright, A.D., et al., *Electron–phonon coupling in hybrid lead halide perovskites.* Nature Communications, 2016. **7**(1): p. 11755.
33. Chen, X., et al., *Excitonic Effects in Methylammonium Lead Halide Perovskites.* The Journal of Physical Chemistry Letters, 2018. **9**(10): p. 2595-2603.
34. Saive, R., *S-Shaped Current–Voltage Characteristics in Solar Cells: A Review.* IEEE Journal of Photovoltaics, 2019. **9**(6): p. 1477-1484.
35. Werner, J., et al., *Improving Low-Bandgap Tin–Lead Perovskite Solar Cells via Contact Engineering and Gas Quench Processing.* ACS Energy Letters, 2020. **5**(4): p. 1215-1223.
36. Afshari, H., *Temperature Dependent Carrier Extraction and the Effects of Excitons on Emission and Photovoltaic Performance in CsFAMAPbIBr.* ACS Applied Materials and Interfaces, 2022.
37. Baranowski, M. and P. Plochocka, *Excitons in Metal-Halide Perovskites.* Advanced Energy Materials, 2020. **10**(26): p. 1903659.
38. Phung, N., et al., *Photoprotection in metal halide perovskites by ionic defect formation.* Joule, 2022.
39. Nguyen, D.-T., et al., *Quantitative experimental assessment of hot carrier-enhanced solar cells at room temperature.* Nature Energy, 2018. **3**(3): p. 236-242.
40. Piyathilaka, H.P., et al., *Hot-carrier dynamics in InAs/AlAsSb multiple-quantum wells.* Scientific Reports, 2021. **11**(1): p. 10483.
41. Jean Rodière, H.L., Amaury Delamarre, Arthur Le Bris, Sana Laribi, et al., *Hot Carrier Solar Cell: From Simulation to Devices*, in *27th European Photovoltaic Solar Energy Conference and Exhibition (EU PVSEC 2012).* 2012: Frankfurt, Germany. p. 89.
42. Jehl, Z., et al. *Hot Carrier Extraction Using Energy Selective Contacts and Feedback On The Remaining Distribution.* in *2018 IEEE 7th World Conference on Photovoltaic Energy Conversion (WCPEC) (A Joint Conference of 45th IEEE PVSC, 28th PVSEC & 34th EU PVSEC).* 2018.
43. Lim, J.W.M., et al., *Spotlight on Hot Carriers in Halide Perovskite Luminescence.* ACS Energy Letters, 2022. **7**(2): p. 749-756.
44. Esmaielpour, H., et al., *Impact of excitation energy on hot carrier properties in InGaAs multi-*

*quantum well structure.* Progress in Photovoltaics: Research and Applications, 2022. **n/a**(n/a).
45. Lim, J.W.M., et al., *Hot Carriers in Halide Perovskites: How Hot Truly?* The Journal of Physical Chemistry Letters, 2020. **11**(7): p. 2743-2750.
46. Esmaielpour, H., et al., *Effect of occupation of the excited states and phonon broadening on the determination of the hot carrier temperature from continuous wave photoluminescence in InGaAsP quantum well absorbers.* Progress in Photovoltaics: Research and Applications, 2017. **25**(9): p. 782-790.
47. Esmaielpour, H., et al., *Hot carrier relaxation and inhibited thermalization in superlattice heterostructures: The potential for phonon management.* Applied Physics Letters, 2021. **118**(21): p. 213902.
48. Esmaielpour, H., et al., *Investigation of the spatial distribution of hot carriers in quantum-well structures via hyperspectral luminescence imaging.* Journal of Applied Physics, 2020. **128**(16): p. 165704.
49. Chen, Z., et al., *Unified theory for light-induced halide segregation in mixed halide perovskites.* Nature Communications, 2021. **12**(1): p. 2687.
50. Brennan, M.C., et al., *Light-Induced Anion Phase Segregation in Mixed Halide Perovskites.* ACS Energy Letters, 2018. **3**(1): p. 204-213.
51. Cho, J. and P.V. Kamat, *Photoinduced Phase Segregation in Mixed Halide Perovskites: Thermodynamic and Kinetic Aspects of Cl–Br Segregation.* Advanced Optical Materials, 2021. **9**(18): p. 2001440.
52. Meloni, S., et al., *Ionic polarization-induced current–voltage hysteresis in $CH_3NH_3PbX_3$ perovskite solar cells.* Nature Communications, 2016. **7**(1): p. 10334.
53. Eames, C., et al., *Ionic transport in hybrid lead iodide perovskite solar cells.* Nature Communications, 2015. **6**(1): p. 7497.
54. Dimmock, J.A.R., et al., *A metallic hot-carrier photovoltaic device.* Semiconductor Science and Technology, 2019. **34**(6): p. 064001.
55. David, K.F., R.W. Vincent, and R.S. Ian, *Pathways to hot carrier solar cells.* Journal of Photonics for Energy, 2022. **12**(2): p. 022204.
56. Dunfield, S.P., et al., *From Defects to Degradation: A Mechanistic Understanding of Degradation in Perovskite Solar Cell Devices and Modules.* Advanced Energy Materials, 2020. **10**(26): p. 1904054.